\documentstyle[osa,manuscript,epsbox]{revtex}  

\begin{document}                

\title{Segregation and Phase Inversion in a Simple Granular System.}

\author{Akinori Awazu\footnote{E-mail: awa@complex.c.u-tokyo.ac.jp}\\ }

\address{Department of Pure and Applied Sciences University of Tokyo,\\
Komaba 3-8-1, Meguro-ku, Tokyo 153-8902, Japan.}

\maketitle
\begin{abstract}
The segregation and the phase inversion are investigated through a 
simple granular system which consists of only two inelastic hard spheres 
in a square box with an energy source. With the variation of 
the coefficient of restitution, the mass ratio between two spheres or 
the box size, we show that two types of segregated states and 
crossover between them are realized in such a small simple system.
\vspace{2mm}
PACS number(s):\\
\end{abstract}

Granular materials exhibit various complex phenomena\cite{re1,re2}. Examples 
are segregations of particles mixtures with different properties 
which appear by shaking or stirring them, or confined them in a horizontally 
rotating drum, and so on
\cite{s1,s11,s12,s2,s9,s10,s23,s4,s5,byt,s27,s6,awa0,Hon,Hon1,Hon2}. 
Recently, many studies for several segregated patterns and the crossover 
between them are reported\cite{s4,s5,byt,s27,s6,awa0,Hon,Hon1,Hon2}.

In this paper, instead of carrying out simulations with many particles, 
we choose a simple system consisting of inelastic hard spheres with 
different masses. Although the system is so simple, it will 
be shown to exhibit segregation of particles. This system is expected as a 
simple model of local processes of highly excited granular mixtures.

Such small systems with elastic hard discs are recently investigated, and 
are found to show a prototype of solid-liquid phase transition, glass 
transition, and the transition between the static and dynamic 
friction\cite{speedy,awa1,awa2}. 
In the following, we show that this simple system realizes the crossover 
between two segregated states like the change between different 
segregated patterns obtained in recent 
studies\cite{s4,s5,byt,s27,s6,awa0,Hon,Hon1,Hon2}.

The system under consideration consists of two inelastic hard sphere  
particles with unit radius and different masses which are confined in a 
two-dimensional square box (Fig. 1). The left-hand wall is set at the 
origin of the horizontal axis and is in contact with the energy source. 
All walls are rigid and have a length larger than $4$. 
The interaction between a particle and the walls without an energy source 
occurs only through hard-core collisions. 
We give the position of a light and a heavy particle in the horizontal 
direction $x_{L}$ and $x_{H}$ as the distance from the left 
wall, and the mass $m_{L}=1$ and $m_{H}=M > 1$. 
The interaction between two particles occur through inelastic hard-core 
collisions with the the coefficient of restitution $e$. 
A particle hitting the left-hand wall in contact with the energy source 
with the velocity $(v_{h},v_{v})$ bounces back with the velocity 
$(V_{h},V_{v})$ ($V_{h}>0$). Here, subscripts
$h$ and $v$ indicate, respectively, the horizontal and vertical 
directions. In this paper, we employ a heat bath of the temperature $T$ 
as the energy source because of the simplicity. Then, the velocity 
$(V_{h},V_{v})$ is chosen randomly from the probability distributions 
$P_{h}(V_{h})$ and $P_{v}(V_{v})$ \cite{hh}:
\begin{equation}
P_{h}(V_{h})=\frac{m_{i} V_{h}}{T} \exp (-\frac{m_{i} V_{h}^{2}}{2 T})
\end{equation}
\begin{equation}
P_{v}(V_{v})=(\frac{m_{i}}{2 \pi T})^{\frac{1}{2}} \exp (-\frac{m_{i} V_{v}^{2}}{2 T}),
\end{equation}
where $T$ is the temperature of the heat bath fixed as unity. 
(We give the Boltzmann constant as $1$.) 

In the following, we perform the simulation of the above system with the 
length of walls $S$ is $=4.05$, $= 4.2$, and $= 4.5$. 
For this range of $S$, particles are densely packed and the 
collision between particles occur frequently. Now, in order to characterize 
the distribution of two particles, we define the segregation parameter 
$\delta  \equiv <x_{L} - x_{H}>_{t}$ where $< >_{t}$ means the time average. 
If $\delta \sim 0$, the particles are not segregated while the sign of 
$\delta$ gives the average configuration of the two particles. 

Now, we draw the phase diagram of the model against $\{M, e \}$ ($M > 1$ 
and $e < 1$), according to the size of $S$. Figure 2 shows the diagram for 
(a) $S=4.05$, (b) $S=4.2$ and (c) $S=4.5$. Here, $\delta > 0$ holds in the 
region with $+$, $\delta < 0$ holds in the region with $-$, and 
$\delta \sim 0$ holds in the shadow area drawn by multiple of points. 
(In this paper, we regard the case with $-(S-d) \times 10^{-3} < \delta < 
(S-d) \times 10^{-3}$ as $\delta \sim 0$ where $d$ is the diameter of the 
particle.)
Independently of $S$, each phase diagram has following characteristics. 
For small $M$, $\delta \sim 0$ or $\delta > 0$ holds over all the range of 
$e < 1$. When $M$ is increased, the system realizes two states with 
$\delta > 0$ and $\delta < 0$ and the crossover between them depends on $e$. 
In particular, critical value $e$ to realize the crossover between states with 
$\delta < 0$ and $\delta > 0$ is independent of $M$ for large 
limit of $M$. However, crossover points shift to larger $e$ and larger $M$ 
with the increase of $S$. 
This means the change of the packing fraction of particles is also relevant 
to the crossover between $\delta < 0$ and $\delta > 0$ states.
Moreover, the crossover points for $S=4.05$ form a curve 
with 'N' shape, while the curve becomes smooth with the increase of $S$ 
like in Fig. 2 (b) and (c).

Figure 3 (a) shows $\delta$ as a function of $e$ foe $M=1.5$, 
$M=2.0$, $M=3.65$ and $M=6.0$ with $S=4.05$. 
For large $M$, $\delta$ takes a minimum value at about $e \sim 0.825$ and 
the minimum value decreases with the increase of $M$.
In Fig. 3 (b), $\delta$ is plotted as a function of $M$ for $e=0.825$, 
$e=0.625$ and $e=0.55$. Here, $\delta$ has a peak at $M \sim 1.3$. 
With the decrease of $e$, $\delta$ for larger $M$ becomes larger than that 
for $M$ at the peak. 

To expalin the existence of the two segregated states with $\delta > 0$ and 
$\delta < 0$, we consider two effects determining 
the sign of $\delta$, respectively; One effect contributes to 
the increase of $\delta$ and the other contributes to the 
decrease of $\delta$ with the decrease of $e$.

First, we study the former effect. The light particle's velocity just after 
the contact with the energy source on the left-hand wall tends to be faster 
than the heavy particle's because these velocities are a decrease function 
of the mass of each particle.On the other hand, by iterations of the 
collision between two particles, their velocities approach with each other 
because $e < 1$. This implies that the light particle tends to be located 
farther from the left-hand wall than to the heavy particle. 
Thus, $\delta$ is expected to increase with the decrease of $e$. 
The contribution of this effect is expected to increase with $1-e$, and 
as a rough approximation, it is assumed to be proportional to $1-e$.

Second, we study the effect by which $\delta$ is decreased with the 
decrease of $e$. If $e$ is given a value close to $1$, the approach of two 
particles' velocities is slow. Then if $M >> 1$, the light 
particle moves much faster than the heavy one for most of the time and the 
collision between two particles occurs frequently. Then, the light particle 
behaves like a potential barrier for the motion of the heavy particle. 
In this case, the heavy particle's motion to go accross this 
potential barrier is important to determine the particle distribution. 
The kinetic energy of the heavy particle with the case of 
$x_{L} < x_{H}$ is smaller than that of $x_{L} > x_{H}$ because, in the 
former case, the heavy particle can not make a contact with the heat bath 
directly and the energy is supplied only by collisions with the light 
particle. This means that the mean velocity of the heavy particle of the 
case $x_{L} < x_{H}$ is smaller than that of $x_{L} > x_{H}$, while the 
ratio between velocities is roughly estimated to be $e : 1$. Then, 
the ratio between the time required to switch from 
$x_{L} < x_{H}$ to $x_{L} > x_{H}$ and that to switch from 
$x_{L} > x_{H}$ to $x_{L} < x_{H}$ is given as $1 : e$. 
Here, this effect on $\delta$ is prominent only for the case with the large 
collision frequency which is almost propotional to $e$.
In addition, this collision frequency increases with the decrease of $S$. 
Then, the contribution from the above effect is approximately 
estimated as $\delta_{1} \sim (e-1)e/C(S)$ (C is an increase function of S.).

By the combination of these effects, $\delta$ for large $M$ is given as 
$\delta \sim  A \delta_{0} + B \delta_{1} = (1-e)(A-\frac{B}{C(S)}e)$.
Here, A and B are given as positive constant values.
With adequate $A$, $B$, and $C$ holding $0 < A C(S)/B < 1$, $\delta$ 
takes a negative value for large $e$ ($1 >  e > A C(S)/B $) and a positive 
value for small $e$ ($0 < e < A C(S)/B $). This result also gives that 
the crossover value of $e$ between the state with $\delta < 0$ and that with 
$\delta > 0$ becomes larger with the increase of $S$.
If $M$ is small, the contribution of $\delta_{1}$ is expected little. 
Thus, $\delta > 0$ is realized for small $M$.

In this paper, the mass segregation and the phase inversion are 
investigated through a system which consists of only two inelastic hard 
spheres in a square box with heat bath. With the variation of the 
coefficient of restitution, the mass ratio or the box size, two types of 
states with different particle distributions and crossover between them 
are observed for the case with a large mass ratio 
between two particles. The system we studied here may look too small and 
simple. However, we expect that this system can describe local dynamics of 
highly excited granular systems, and provide a basis for the understanding 
of particle segregations in granular system consisting of many particles. 

The effects of the gravity, the size differences 
and the friction should be considered in order to investigate the generality 
of obtained phenomena like the crossover between different segregated 
states\cite{s4,s5,byt,s27,s6,awa0,Hon,Hon1,Hon2}.  
Further analytical study of this system to clarify the presented 
behavior as well as the study of systems with three or more particles are 
necessary in future.

The author is grateful to K.Kaneko, M. Mizuguchi, H. Hayakawa, 
H. Nishimori and M. Otsuki for useful discussions. This research was
supported in part by Grant-in-Aid for JSPS Felows 10376.

\newpage

\begin{figure}[h]
\center
\caption[]{Illustration of the system with two inelastic hard spheres and 
energy source (left).}
\end{figure}

\begin{figure}[h]
\center
\caption[]{Phase diagrams of the system for each set $(M , e)$ with 
(a) $S = 4.05$, (b) $S = 4.2$ and (c) $S = 4.2$. $+$ and $-$ mean the 
sign of $\delta$, and $\delta \sim 0$ in the shadow area. }
\end{figure}

\begin{figure}[h]
\center
\caption[]{(a) $\delta$ as the function of $e$ 
for $M=1.5$, $M=2.0$, $M=3.65$ and $M=6.0$, and (b) $\delta$ as the 
function of $M$ for $e=0.825$, $e=0.625$ and $e=0.55$ with $S=4.05$.}
\end{figure}


\begin{thebibliography}{999}

\bibitem{re1}
H. Hayakawa, H. Nishimori, S. Sasa and Y-h. Taguti,
Jpn. J. Appl. Phys. {\bf34} 397 (1995). 

\bibitem{re2}
H. M. Jeager and S. R. Nagel, Science {\bf255} 1523 (1990)

\bibitem{s1}
C. Williams, Powder Technol. {\bf 15} 254 (1976).

\bibitem{s11}
A. Rosato, K. J. Strandburg, F. Prinz and R. H. Swendsen, Phys. Rev. Lett. {\bf 58}, 1038 (1987).

\bibitem{s12}
R. Jullien and P. Meakin, Phys. Rev. Lett {\bf 69}, 640 (1992).

\bibitem{s2}
J. Duran, J. Rajchenbach and E. Clement, Phys. Rev. Lett {\bf 70}, 2431 (1993).

\bibitem{s9}
O. Zik, D. Levine, S. G. Lipson, S. Shtrikman and
J. Stavans, Phys. Rev. Lett. {\bf 73}, 644 (1994).

\bibitem{s10}
K. M. Hill, A. Caprihan and J. Kakalios, Phys. Rev. Lett. {\bf
78}, 50 (1997).

\bibitem{s23}
M.Nakagawa, NATO ASI Series, Series E: Applied Siences - 
Vol.350 Physics of Dry Granular Media (H.J.Herrmann, J.-P.Hovi, and
S.Luding) 703 (1998).

\bibitem{s4}
J. B. Knight, H. M. Jaeger and S.R.Nagel, Phys. Rev. Lett. {\bf 70}, 3278 (1993).

\bibitem{s5}
T. Shinbrot  and F. J. Muzzio, Phys. Rev. Lett {\bf 81}, 4365 (1998).

\bibitem{byt}J H. Bytnar, J. O. G Parent, H. Henein and J. Iyengar,
International Journal of Powder Metallugy. {\bf 31}, 37 (1995).

\bibitem{s27} 
J. L. Turner and M.Nakagawa, Powder Technol. {\bf 113}, 119 (2000)

\bibitem{s6}
H A. Makse, P. Cizeau and H. E. Stanley, Phys. Rev. Lett. {\bf 78}, 3298
(1997).

\bibitem{awa0}
A. Awazu, Phys. Rev. Lett. {\bf 80}, 4585 (2000).

\bibitem{Hon}
D. C. Hong and P. V. Quinn, Phys. Rev. Lett. {\bf 86}, 3423 (2001).
	
\bibitem{Hon1}
J. A. Both and D. C. Hong, Phys. Rev. Lett. {\bf 88}, 124301 (2002).

\bibitem{Hon2}
M. Nicodemi, A. Fierro and A Coniglio, Europhys. Lett. {\bf 60}, 684 (2002).

\bibitem{speedy}
R. J. Speedy Physica A {\bf 210}, (1994) 341.

\bibitem{awa1}
A. Awazu, Phys. Rev. E. {\bf 63}, 032102 (2001).

\bibitem{awa2}
A. Awazu, J. Phys. Soc. Jpn {\bf 71}, 15 (2002).

\bibitem{hh}
R. Tehver, F. Toigo, J. Koplik and J. R. Banavar, Phys. Rev. {\bf E57}, 57 (1998). 
\end{thebibliography}
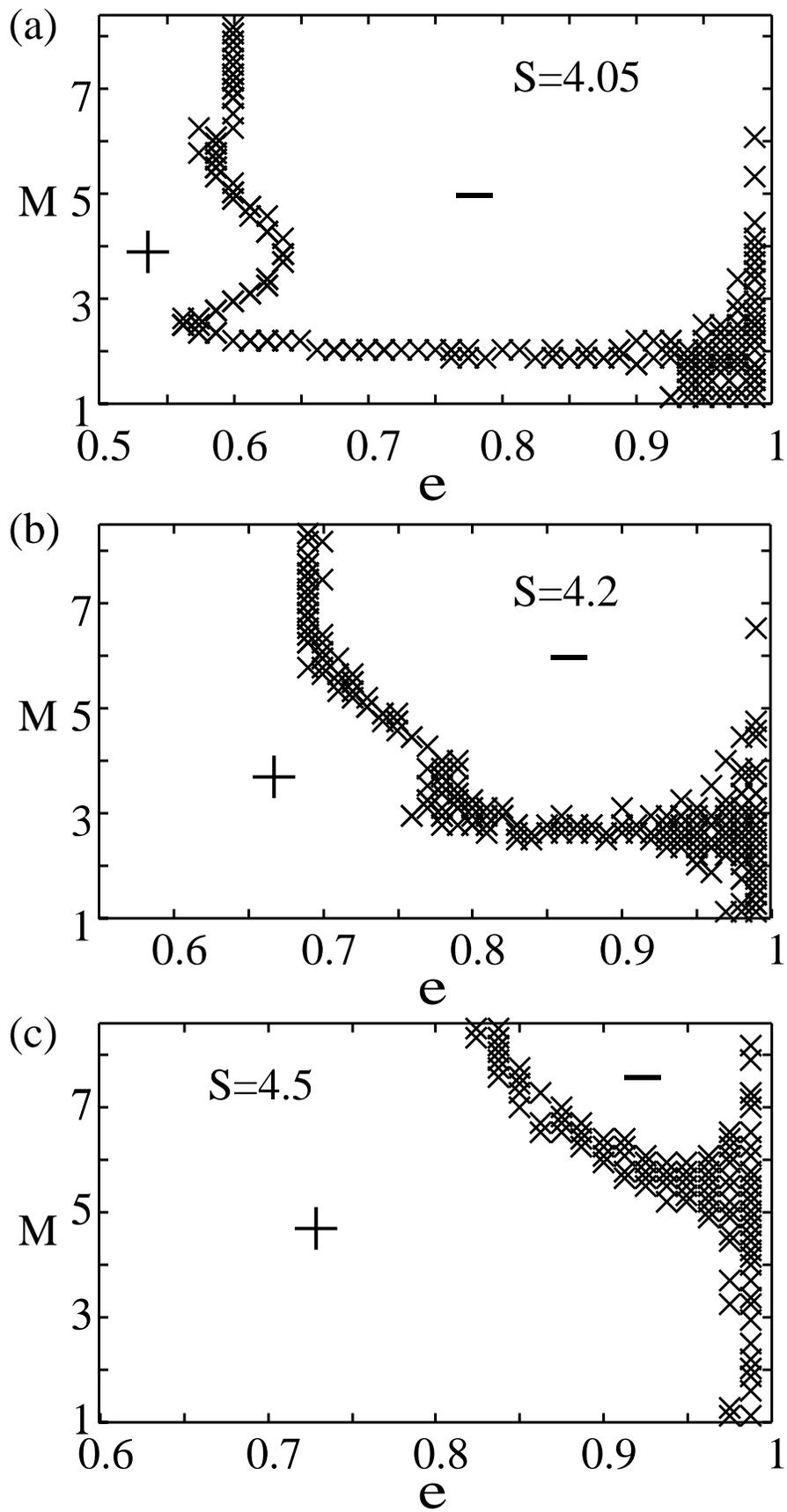
\end{document}